\documentclass[submission, Phys]{SciPost}

\hypersetup{
    colorlinks,
    linkcolor={red!50!black},
    citecolor={blue!50!black},
    urlcolor={blue!80!black}
}

\usepackage[bitstream-charter]{mathdesign}
\DeclareSymbolFont{usualmathcal}{OMS}{cmsy}{m}{n}
\DeclareSymbolFontAlphabet{\mathcal}{usualmathcal}

\begin{document}

\begin{center}{\Large \textbf{
Search for charged lepton flavor violation in $J/\psi$ decays at BESIII\\
}}\end{center}

\begin{center}
Xudong Yu\textsuperscript{1$\star$}
\end{center}

\begin{center}
{\bf 1} Peking University
\\
* yuxd@stu.pku.edu.cn
\end{center}

\begin{center}
\today
\end{center}


\definecolor{palegray}{gray}{0.95}
\begin{center}
\colorbox{palegray}{
  \begin{tabular}{rr}
  \begin{minipage}{0.1\textwidth}
    \includegraphics[width=30mm]{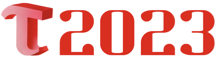}
  \end{minipage}
  &
  \begin{minipage}{0.81\textwidth}
    \begin{center}
    {\it The 17th International Workshop on Tau Lepton Physics}\\
    {\it Louisville, USA, 4-8 December 2023} \\
    \doi{10.21468/SciPostPhysProc.?}\\
    \end{center}
  \end{minipage}
\end{tabular}
}
\end{center}

\section*{Abstract}
{\bf
In the Standard Model, charged lepton flavor violation (CLFV) is heavily suppressed by tiny neutrino mass, while many theoretical models can enhance CLFV effects up to a detectable level. The observation of any CLFV process would be a clear signal of new physics beyond SM. BESIII experiment collected 10 billion $J/\psi$ data and searched for CLFV processes $J/\psi\to e\tau$ and $e\mu$. The upper limits at the 90\% confidence level are determined to be $\mathcal{B}(J/\psi\to e\tau)<7.5\times 10^{-8}$ and $\mathcal{B}(J/\psi\to e\mu)<4.5\times 10^{-9}$, respectively. Improving the previously experimental limits by two orders of magnitudes, the results are the most stringent CLFV searches in the heavy quarkonium system.
}

\vspace{10pt}
\noindent\rule{\textwidth}{1pt}
\tableofcontents\thispagestyle{fancy}
\noindent\rule{\textwidth}{1pt}
\vspace{10pt}

\section{Introduction}
\label{sec:intro}
In the standard model (SM), the flavor number is not a symmetry of a Lagrangian-like charge. In the quark scenario, the quark flavor number is violated in weak decays according to Cabibbo-Kobayashi-Maskawa matrix, while in the lepton case, the lepton flavor number should also be violated according to Pontecorvo-Maki-Nakagawa-Sakata Matrix. Three neutral leptons, $i.e.,$ neutrinos, are found to oscillate and mix among themselves. However, flavor mixing among charged leptons has never been observed so far, which remains a mystery to be explored~\cite{Li:2022hgy}.

According to Glashow–Iliopoulos–Maiani mechanism, the Charged Lepton Flavor Violation (CLFV) event rate is too small to be observed if the only CLFV source is from non-zero tiny neutrino mass. For example, the branching fraction (BF) of $\mu\to e\gamma$ is suppressed below $10^{-54}$. Various theoretical models beyond the SM can greatly enhance the CLFV effects. For example, two Higgs doublet model with extra Yukawa couplings~\cite{Branco:2011iw, Crivellin:2013wna, Hou:2021zqq}, a model that constitutes a simple example of tree-level off-diagonal Majorana couplings not suppressed by neutrino masses~\cite{Escribano:2021uhf}, Supersymmetry (SUSY)-based grand unified theories (GUT)~\cite{Dimopoulos:1981zb}, SUSY with vector-like leptons~\cite{Kitano:2000zw} and a right-handed neutrino~\cite{Borzumati:1986qx}, the minimal SUSY model with gauged baryon number and lepton number~\cite{Dong:2017ipa}, models with a $Z^{\prime}$~\cite{Bernabeu:1993ta}, etc. Therefore, the observation of any CLFV process would undoubtedly be a clear signal of new physics (NP) beyond the SM.

Search for CLFV processes is one of the most concerned directions in the field of particle physics. Experimental measurements can be roughly divided into two aspects, one is performed in dedicated muon beams, and the other is conducted in colliders. The upper limit (UL) of $\mu^+\to e^+\gamma$ process is determined to be $4.2\times10^{-13}$ at 90\% confidence level (CL) by MEG experiment at PSI~\cite{MEG:2016leq}, and MEG II will further improve it. The $\mu e\gamma$ dipole can also be probed by $\mu\to3e$ in Mu3e and $\mu N\to eN$ by COMET and Mu2e in the future. Muon beams give the most powerful limits but cannot study $\tau$, $Z$, Higgs bosons, pseudoscalar mesons ($K,\pi, B$), and vector mesons ($\phi, J/\psi, \Upsilon$) decays. For the CLFV decays of $J/\psi$, the decay rates are predicted up to a detectable level of around $10^{-8}\sim10^{-16}$ using model-independent methods~\cite{Nussinov:2000nm, Gutsche:2011bi}, unparticle physics~\cite{Sun:2012hb}, and the minimal supersymmetric model with gauged baryon number and
lepton number~\cite{Dong:2017ipa}, etc. With $58\times10^{6}$ $J/\psi$ events, the BES Collaboration obtained ULs of various decays of charmonium states, e.g. $\mathcal{B}(J/\psi\to e\mu)<1.1\times10^{-6}$~\cite{BES:2003zru}, $\mathcal{B}(J/\psi\to e\tau)<8.3\times10^{-6}$, and $\mathcal{B}(J/\psi\to \mu\tau)<2.0\times10^{-6}$~\cite{BES:2004jiw}. Based on $225\times10^{6}$ $J/\psi$ events, an UL of $\mathcal{B}(J/\psi\to e\mu)<1.6\times10^{-7}$ was obtained by the BESIII Collaboration~\cite{BESIII:2013jau}. This paper summarized two recent published results of searching for $J/\psi\to e^{\pm}\tau^{\mp}$ with $\tau^{\mp}\to\pi^{\mp}\pi^0\nu_\tau$ based on $10 \times 10^9$ $J/\psi$ events and $J/\psi\to e^{\pm}\mu^{\mp}$ using $8.998\times10^{9}$ $J/\psi$ events collected with the BESIII detector~\cite{BESIII:2021slj,BESIII:2022exh}.

\section{BESIII detector}
The BESIII detector~\cite{Ablikim:2009aa} consists of four sub-detectors, i.e., multilayer drift chamber~(MDC), time-of-flight system (TOF), electromagnetic calorimeter (EMC), and muon counter (MUC). BESIII operates in the tau-charm energy region with a center-of-mass from 2.0 to 4.95~GeV, with a peak luminosity of $1.1 \times10^{33}\;\text{cm}^{-2}\text{s}^{-1}$.

\section{CLFV results from BESIII}
\label{sec:clfv_besiii}
The analyses are performed based on $J/\psi$ events collected in 2009, 2012, 2018, and 2019 at $\sqrt{s}=3.097$ GeV ($J/\psi\to e\mu$ without data in 2012)~\cite{BESIII:2021cxx}. To avoid potential bias, a semi-blind analysis approach is employed, where 10\% of the complete data sample is randomly selected for analysis.

\subsection{Search for CLFV decay $J/\psi \to e\tau$}
The final-state $e$ in the decay $J/\psi\to e\tau$ exhibits monochromatic behavior, hence the momentum $P_e$ and the recoiling mass $M_{e\_\mathrm{recoil}}$ must fall within $1.009$ GeV/$c$ $<P_e<$ 1.068 GeV/$c$ and $1.742$ GeV/$c^2$ $<M_{e\_\mathrm{recoil}}<1.811$ GeV/$c^2$, respectively. The $\tau$ lepton is reconstructed by $\pi\pi^0\nu_\tau$ channel. Since the neutrino cannot be detected in the BESIII detector, the signal is searched in $U_{\mathrm{miss}}$ distribution, where $U_{\mathrm{miss}}$ is defined as $U_{\mathrm{miss}}=E_{\mathrm{miss}} - c|\vec{P}_{\mathrm{miss}}|$. The missing energy $E_{\mathrm{miss}}$ is calculated as $E_{\mathrm{miss}} = E_{\mathrm{CMS}} - E_{e} - E_{\pi} - E_{\pi^0}$, where $E_{\mathrm{CMS}}$ represents the center-of-mass energy of the initial $e^+e^-$ system, and $E_e, E_\pi$ and $E_{\pi^0}$ denote the energies of the electron, the charged pion and the neutral pion in the rest frame of the $e^+e^-$ system. To effectively suppress the background events, the missing energy should exceed 0.43 GeV. The missing momentum $\vec{P}_{\mathrm{miss}}$ is determined by $\vec{P}_{J/\psi} - \vec{P}_{e} - \vec{P}_{\pi} - \vec{P}_{\pi^0}$, where $\vec{P}$ are the corresponding momenta in the rest frame of $e^+e^-$ system. The $U_{\mathrm{miss}}$ signal region is defined as $-0.081$ GeV $<U_{\mathrm{miss}}<$ $0.112$ GeV. In data sample I (II), collected in 2009 and 2012 (2018 and 2019), there are 13 (69) candidate events observed. The detection efficiency for signal sample I (II) is
determined to be $(20.24 \pm 0.05)\%$ ($(19.37 \pm 0.02)\%$).

The primary background contamination arises from continuum processes, such as radiative Bhabha scattering and hadronic $J/\psi$ decays like $J/\psi \to \pi^+\pi^-\pi^0$. The continuum background is analyzed using control samples at $\sqrt{s}=3.08$ GeV and 3.773 GeV. The normalized background events originating from the continuum processes are estimated assuming a $1/s$ dependence of the cross section. The expected continuum background events are calculated to be $5.8\pm1.8$ ($37.9\pm11.5$) for data sample I (II). The background from $J/\psi$ decays is investigated using the inclusive MC samples. The estimated normalized background events from the $J/\psi$ decays are $1.1 \pm 0.8$ ($25.7 \pm 6.4$) for data sample I (II).

\begin{figure}[h]
\centering
\includegraphics[width=0.5\textwidth]{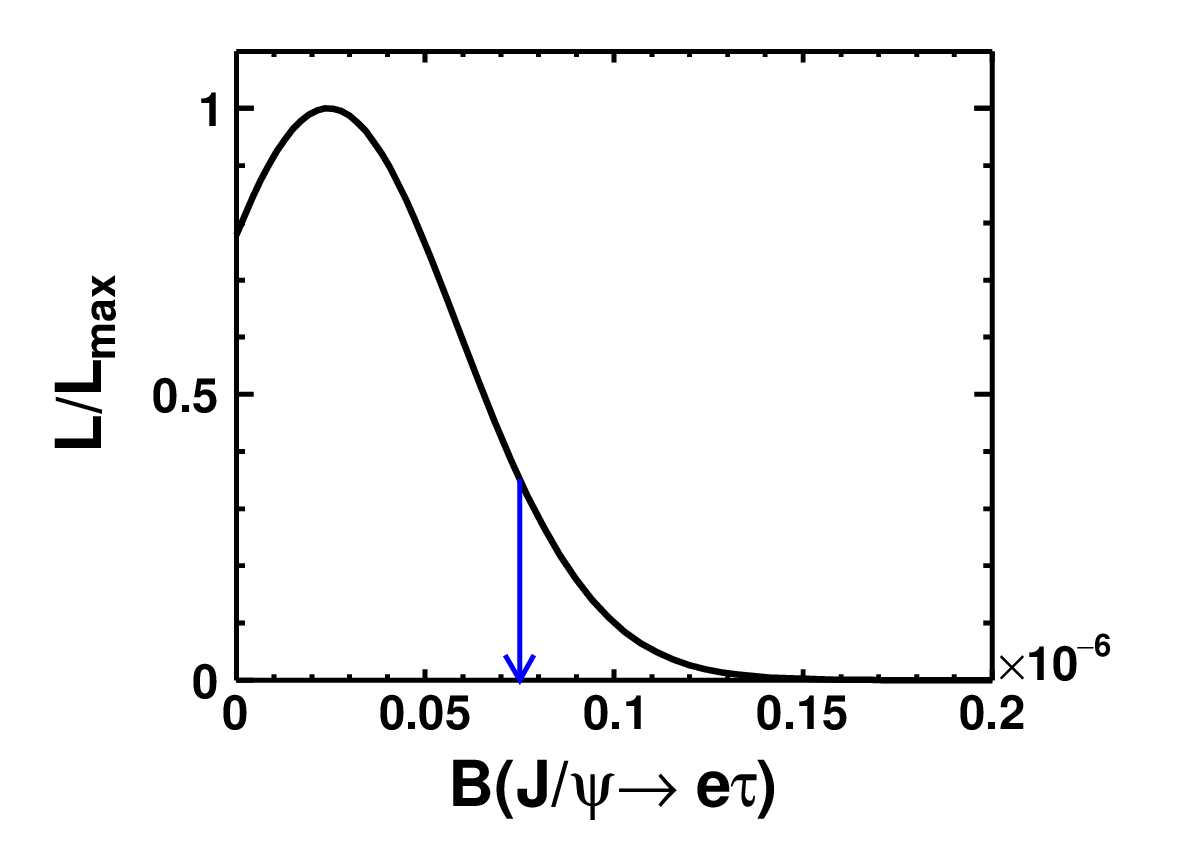}
\caption{The combined likelihood distribution as a function of the BF of the data samples. The arrow points to the position of the UL at 90\% CL. Figure adapted from Ref.~\cite{BESIII:2021slj}}
\label{fig:Jpsi_etau_UL}
\end{figure}

As no significant signal is observed, a maximum likelihood estimator, extended from the profile-likelihood approach, is utilized to determine the UL on the BF of $J/\psi\to e\tau$. The UL is determined to be $\mathcal{B}(J/\psi\to e\tau)<7.5\times 10^{-8}$ at the 90\% CL, as shown in Figure~\ref{fig:Jpsi_etau_UL}. The obtained result enhances the previous best limits by two orders of magnitude and is comparable with the theoretical predictions.

\subsection{Search for CLFV decay $J/\psi \to e\mu$}
Two back-to-back oppositely charged tracks reconstruct each $J/\psi$ candidate. To mitigate cosmic ray interference, events are filtered based on a TOF timing difference requirement of less than 1.0~ns. Furthermore, the acollinearity angle, defined as $|\Delta\theta|=|180^{\circ}-(\theta_1+\theta_2)|$, is required to be less than $1.2^{\circ}$, where $\theta_1$ and $\theta_2$ represent the polar angles of the two tracks. The coplanarity angle, denoted as $|\Delta\phi|=|180^{\circ}-|\phi_1-\phi_2||$, must less than $1.5^{\circ}$, with $\phi_1$ and $\phi_2$ being the corresponding azimuthal angles. Energy loss $\mathrm{d}E/\mathrm{d}x$ in MDC, deposited energy in EMC and hit information in MUC are utilized to identify electrons and muons. The signal region is defined based on the principles of energy and momentum conservation, requiring $\sum\vec{p}/\sqrt{s}\le0.02$ and $0.95\le E_{\mathrm{vis}}/\sqrt{s}\le 1.04$, where $\sum\vec{p}$ represents the magnitude of the vector
sum of the momenta, and $E_{\mathrm{vis}}$ denotes the total reconstructed energy of the electron and muon in the event. The detection efficiency for the signal process is determined to be ($21.18 \pm 0.13\%$), with 29 candidate events observed within the signal region.

Two types of background events contaminate the signal region. One is continuum background, such as $e^+e^-\to e^+e^-(\gamma), \mu^+\mu^-(\gamma)$. The other type is $J/\psi$ decaying into $e^+e^-$, $\mu^+\mu^-$, $\pi^+\pi^-$, $K^+K^-$, $p\bar{p}$. Control samples at $sqrt{s}=$ 3.773 GeV, 3.510 GeV, and 3.090 GeV are used for continuum background. By assuming a $1/s$ energy-dependence of the cross sections, the normalized number of continuum backgrounds at the $J/\psi$ peak, $N^{\mathrm{norm},k}_{\mathrm{bkg2}}$, is estimated to be $12.0\pm3.7$. The contribution of the $J/\psi$ decay background is analyzed using inclusive and exclusive MC samples. It is estimated to be $N^{\mathrm{norm}}_{\mathrm{bkg1}}=24.8\pm1.5$.

\begin{figure}[h]
\centering
\includegraphics[width=0.5\textwidth]{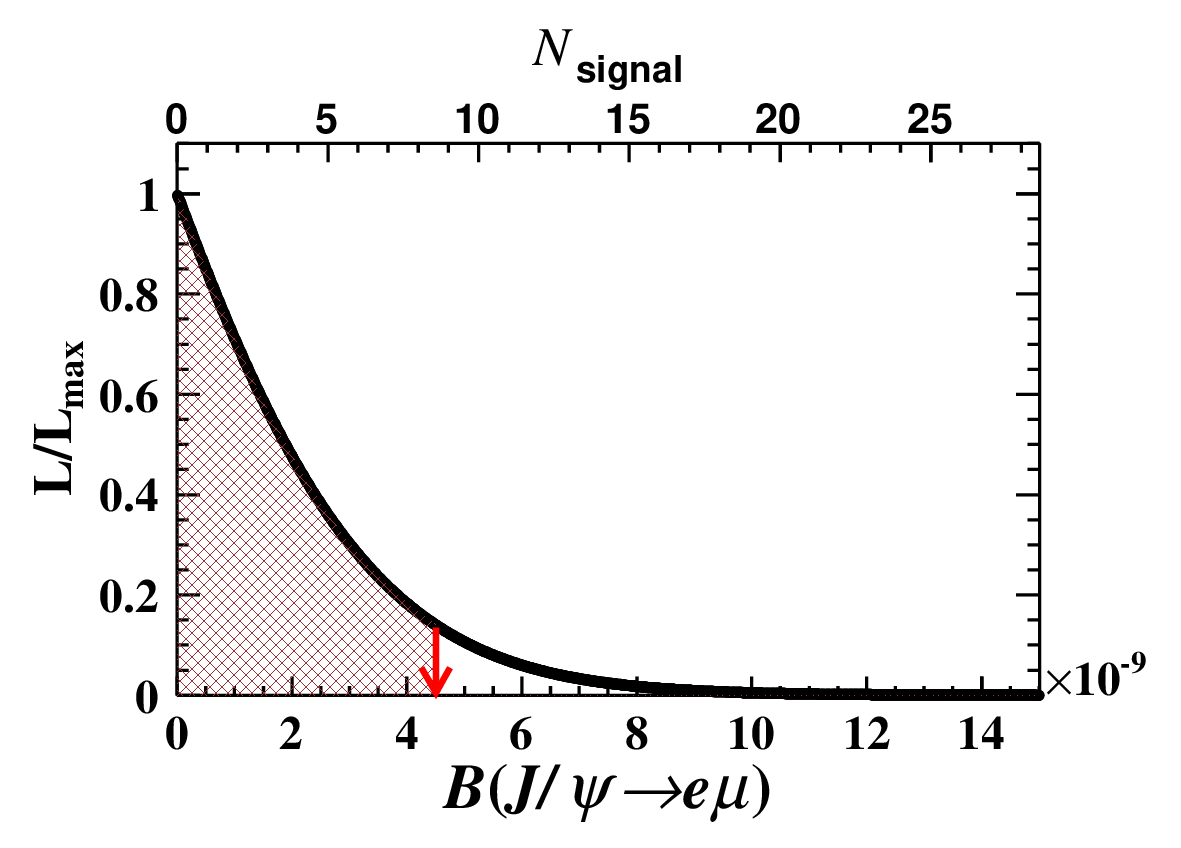}
\caption{Normalized likelihood distribution as a function of the assumed $\mathrm{B}(J/\psi \to e\mu)$. The red arrow points to the position of the UL at the 90\% CL. The upper x-axis shows the corresponding values of the signal yields. Figure adapted from Ref.~\cite{BESIII:2022exh}}
\label{fig:Jpsi_emu_UL}
\end{figure}

By employing the profile likelihood method, the UL on the BF is established as $\mathrm{B}(J/\psi\to e\mu)<4.5\times10^{-9}$ at the 90\% CL, as depicted in Figure~\ref{fig:Jpsi_emu_UL}. This result enhances the previous limit by more than 30 times, making it by far the most stringent constraint on CLFV within heavy quarkonium systems. Furthermore, this limit constrains the parameter spaces of new physics models.

\section{Summary}
BESIII has collected the largest data samples of $J/\psi$ on threshold in the world. The high statistics $J/\psi$ data offers a distinctive opportunity for an extensive exploration of CLFV searching for new physics beyond the SM. This proceeding summarizes the studies of the CLFV processes $J/\psi\to e\tau$ and $J/\psi\to e\mu$ at BESIII. The ULs are determined to be $\mathcal{B}(J/\psi\to e\tau) < 7.5 \times 10^{-8}$ and $\mathcal{B}(J/\psi\to e\mu) < 4.5 \times 10^{-9}$ at the 90\% CL.

\bibliography{SciPost_Example_BiBTeX_File.bib}

\nolinenumbers

\end{document}